\documentclass[12pt,preprint]{aastex}
\usepackage{subfigure}

\def \ts{\langle\theta_{s}^2\rangle}

\IfFileExists{srcltx.sty}{\usepackage[active]{srcltx}}

\shorttitle{TIME STRUCTURE}
\shortauthors{PROSEKIN et al.}

\begin{document}

\title{Time structure of gamma-ray signals generated in line-of-sight interactions of cosmic rays from distant blazars}

\author{Anton Prosekin\altaffilmark{1,6}, Warren~Essey\altaffilmark{2},  Alexander Kusenko \altaffilmark{3,4}, and Felix~Aharonian\altaffilmark{5,1}}

\altaffiltext{1}{Max Planck Institute, Heidelberg, Germany}
\altaffiltext{2}{ICCS, University of California, Berkeley,  CA 94708-1003, USA}
\altaffiltext{3}{Department of Physics and Astronomy, University of California, Los Angeles, CA 90095-1547, USA}
\altaffiltext{4}{Kavli Institute for the Physics and Mathematics of the Universe,
University of Tokyo, Kashiwa, Chiba 277-8568, Japan}
\altaffiltext{5}{School of Cosmic Physics, Dublin Institute for Advanced Studies, 31 Fitzwilliam Place,
Dublin 2, Ireland}
\altaffiltext{6}{Fellow of the International Max Planck Research School for Astronomy and
Cosmic Physics at the University of Heidelberg (IMPRS-HD)}

\begin{abstract}
Blazars are expected to produce both gamma rays and cosmic rays. Therefore, observed high-energy gamma 
rays from distant blazars may contain a significant contribution from 
secondary gamma rays produced along the line of sight by the interactions of cosmic-ray protons 
with background photons.  Unlike the standard models of blazars that consider only the primary photons emitted at the source,
models which include the cosmic-ray contribution predict that even $\sim 10$~TeV photons should be detectable from distant objects 
with redshifts as high as $z \geq 0.1$. Secondary photons contribute to signals of point sources only  if the intergalactic magnetic fields are very small,  
$B \stackrel{\scriptstyle _<}{\scriptstyle _\sim} 10^{-14}$~G, and their detection can be used to set upper bounds on magnetic fields along the line of sight. 
Secondary gamma rays have distinct spectral and temporal features.  We explore the temporal properties of such signals using a semi-analytical formalism 
and detailed numerical simulations, which account for all the relevant processes, including magnetic deflections.  
In particular, we elucidate the interplay of time delays coming from the proton deflections and from the 
electromagnetic cascade, and we find that, at multi-TeV energies,  secondary gamma-rays can show variability on timescales of years for $B \sim 10^{-15}$~G.  
\end{abstract}

\keywords{gamma rays, cosmic rays, active galaxies}

\section{Introduction}

Active Galactic Nuclei (AGN) are expected to be sources of  both cosmic rays and gamma rays.  
While gamma rays have been observed from a number of blazars, the identification of cosmic rays with their sources is impossible (except at the highest energies), because the Galactic magnetic fields change their directions considerably.  However, as long as the intergalactic magnetic fields are relatively small, cosmic rays produced in blazars can travel close to the line of sight and produce secondary gamma rays which would significantly contribute to the radiation observed from the direction of the point sources.  For nearby blazars such contributions are expected to be  small in comparison with direct gamma-ray signals reaching Earth.  However, for more distant blazars, the line-of-sight produced gamma rays can dominate over the direct gamma rays from the source~\citep{Essey:2009zg}.  The   transition occurs because the primary gamma rays are filtered out in their interactions with extragalactic background light (EBL), while the fraction of secondary gamma rays produced by cosmic rays in intergalactic space grows with distance traveled.  Based on the spectra of individual blazars ~\citep{Essey:2009zg,Essey:2009ju,Essey:2010er} and on the trend in spectral softening~\citep{Essey:2011wv}, one expects the secondary contribution to be important for redshifts $z>0.15$ and energies $E>1$~TeV. 

The  intrinsic gamma-ray spectra of some  blazars, after correction for absorption in EBL,  
appear extremely hard, challenging the standard, e.g. the  synchrotron-self-Compton (SSC) or External Compton models  of blazars. 
Several solutions to this problem have been proposed.   Intergalactic cascading of gamma rays from 
blazars in the case of very weak intergalactic magnetic fields (IGMFs) can increase the effective 
mean free path of gamma-rays \citep{Aharonian2002,Taylor2011}, however, for distant blazars 
this effect alone appears to be insufficient to explain the gamma-ray spectra above 1~TeV.
The very hard spectra of primary gamma rays \citep{Katarzynski,Stecker:2007jq,Lefa}, or special features in the sources \citep{2008MNRAS.387.1206A} can help
reconcile the data with theoretical predictions, at the cost of introducing some {\em ad hoc} assumptions. Hypothetical new particles~\citep{2007PhRvD..76l1301D,2012JCAP...02..033H} and Lorentz invariance violation \citep{Protheroe:2000hp} have been invoked to explain the data.

\begin{figure}
\begin{center}
\includegraphics[width=1\textwidth,angle=0]{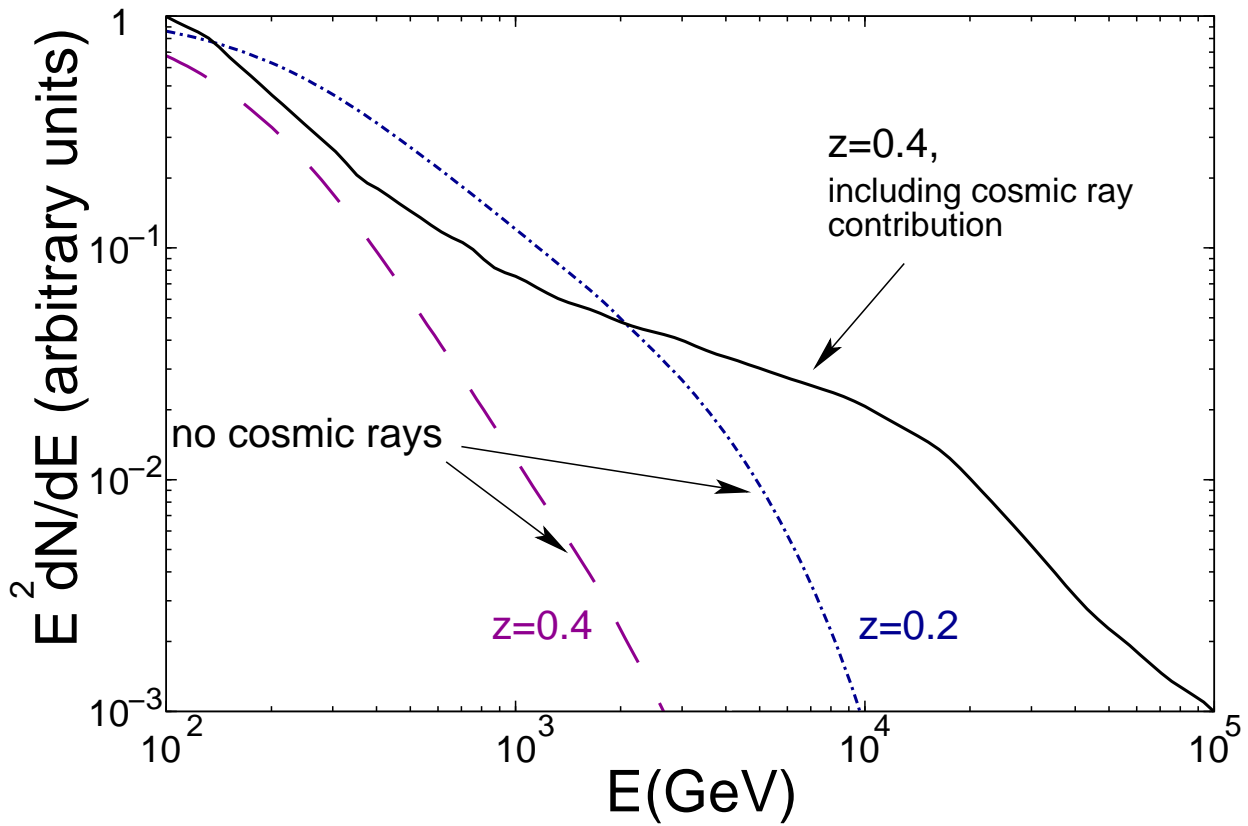}
\caption{\label{fig:z}  Secondary gamma rays produced in line-of-sight interactions of cosmic rays result in harder spectra for distant sources.  
Since most of the observed photons are produced relatively close to the observer, there is less attenuation due to the interactions with EBL.   
}
\end{center}
\end{figure}

The inclusion of cosmic-ray contribution offers an alternative solution to the problem.  Indeed, since a significant fraction of gamma rays in this model 
is produced relatively close to the observer, this model reduces dramatically the impact of gamma-ray 
absorption in EBL. This is illustrated in Fig.~\ref{fig:z}.  Protons with energy $E \geq 10^{16}$~eV propagating through weak IGMFs without 
strong deviations from the line of sight can carry energy from the source close to the observer and can generate a substantial gamma-ray flux 
at multi-TeV energies.  Remarkably, the predicted spectra of secondary  gamma rays depend only on the source redshift (determined from independent observations).
For each source, the power emitted in cosmic rays is the only fitting parameter which can be used to fit the data. As long as the source redshifts are known, 
the predictions of this model are solid and robust.  The spectra calculated for the redshifts of all observed distant blazars provide a very 
good fit to observational data~\citep{Essey:2009zg,Essey:2009ju,Essey:2010er,2011arXiv1107.5576M}, 
with a reasonable  required luminosities in cosmic rays assuming that the escape of 
protons from the source is strongly beamed toward the observer~\citep{Essey:2010er,2012ApJ...745..196R}. 

Confirmation of this model by future observations will have several important  consequences. It will imply that (i) cosmic rays are, indeed, accelerated in AGN, as has long been suspected, but never before confirmed by observations; (ii) intergalactic magnetic fields are fairly small, of the order of several femtogauss ($10^{-15}$~G) or lower~\citep{Essey:2010nd}; and (iii) the problem of intergalactic gamma-rays can be somewhat relaxed, and consequently the upper limits on EBL derived while neglecting the cosmic-ray contributions may need be revised.  Within this model, the expected temporal structure of signals from distant blazars at the highest energies should reflect the time delays cosmic rays undergo in the intergalactic magnetic fields.   We note that while time variability has been observed for {\em nearby} TeV blazars at TeV energies and for distant TeV blazars at energies above a few hundred GeV, no variability has been  reported  so far for {\em distant} TeV blazars at {\em TeV  energies}.  
Here we call {\em distant} those blazars that have large enough redshifts for the primary TeV gamma-rays to die out.  
Based on the spectral fits~\citep{Essey:2009zg,Essey:2009ju,Essey:2010er,2011arXiv1107.5576M},  and the spectral softening of blazars~\cite{Essey:2011wv}, 
one concludes that most blazars with redshift $z\ge 0.15$ should fall in this category.  Since the ratio of gamma-ray luminosity to cosmic-ray luminosity 
can vary from source to source, one expects a mixed population to exist at some intermediate redshifts $0.1<z<0.15$, where stronger cosmic ray emitters 
should be observed in secondary gamma rays, while stronger gamma ray emitters should be observed in primary gamma rays.  Furthermore, if stronger 
IGMFs exist in the directions of specific sources, the secondary contribution can be suppressed.  For example, PKS 2155-304 at $z=0.12$ is an example 
of a source at an intermediate redshift from which primary signals are observed, as indicated by its TeV variability~\cite{2010A&A...520A..83H}. 
Whether the lack of TeV variability is a generic feature of distant blazars, or merely an artifact of low statistics in multi-TeV photons, should be 
clarified in future observations.  An  important issue in this context is the knowledge of the spectral and temporal features of the radiation predicted by the model. 
The spectral features of the radiation have been studied in detail by \cite{Essey:2010er}.

In this paper we consider the extent to which the time variability at high energies should be erased by the cosmic ray propagation delays. We will focus on calculating the Green's function, which corresponds 
to a time delay from an infinitely narrow pulse of protons at the source.  Realistic time profiles can be obtained by convolving the time-dependent source luminosity with this Green's fucntion. However, since a fair fraction of the blazar flaring activity occurs on the time scales much sorter than those we discuss below, in many cases the Green's function can be interpreted as a distribution of photon arrival times from a flaring source.  In applications of our method to data analysis, one can employ time-dependent templates inferred from lower energies.

\section{Basic estimates and scaling laws}

There is no doubt that, for nearby blazars, the primary gamma rays produced at the source are responsible for most of the observed radiation. 
While these objects can also produce cosmic rays, the contribution of secondary gamma rays is not expected to dominate.  
However, for larger distances, the primary gamma-ray component is filtered out above a TeV, while the secondary contribution is enhanced.  
Indeed, the scaling of the primary gamma rays with distance is determined by the losses due to pair production in gamma-ray interactions with extragalactic background light (EBL): 
\begin{eqnarray}
F_{\rm primary,\gamma}(d)& \propto &  \frac{1}{d^2}\exp(-d/\lambda_\gamma). \label{exponential} 
\end{eqnarray}
In contrast, gamma rays generated in line-of-sight interactions of cosmic rays exhibit a very different scaling with distance~\citep{Essey:2009zg,Essey:2009ju,Essey:2010er}: 
\begin{eqnarray}
F_{\rm secondary,\gamma}(d)& \propto &  \frac{\lambda_\gamma}{d^2}\Big(1-e^{-d/\lambda_\gamma}\Big) \nonumber \\
& \sim &  \left \{ 
\begin{array}{ll}
1/d, & {\rm for} \ d \ll \lambda_\gamma, \\ 
1/d^2, & {\rm for} \ d\gg \lambda_\gamma .
\end{array} \right.  \label{line-of-sight} 
\end{eqnarray}
Here $\lambda_\gamma$ is the distance at which EBL opacity to TeV gamma rays is  of the order of $1$. The lack of suppression is due to the fact that the photon backgrounds (CMB and EBL) act 
as a target on which gamma rays are produced by the cosmic rays.  Hence, a higher column density of background photons for a more distant source boosts, not hinders the gamma-ray production. 

As long as IGMFs are weak enough to cause only small deflections, for a sufficiently distant source, secondary gamma rays dominate because they don't suffer from exponential suppression as in Eq.~(\ref{exponential}), which is absent from Eq.~(\ref{line-of-sight}).  The transition from primary to secondary photons occurs when the optical depth to pair production exceeds 1.  The corresponding redshift can also be inferred from the spectral softening of the spectra~\citep{Essey:2011wv}.  Based on these estimates, one can expect the secondary gamma rays to dominate for sources at redshifts 
$z\stackrel{\scriptstyle _>}{\scriptstyle _\sim}0.15 $.  

The success of the spectral fits to the data for secondary gamma rays~\citep{Essey:2009zg,Essey:2009ju,Essey:2010er} can be interpreted as possible   evidence of cosmic ray acceleration in blazars.  Within this interpretation, the beamed energy output in $E>10^{17}$~eV cosmic rays required to fit the observed spectra of distant blazars is of the order of $10^{43}$~erg, or $10^{45}$~erg isotropic equivalent \citep{Essey:2009zg,Essey:2009ju,Essey:2010er}, which is consistent with many models~\citep{2006PhRvD..74d3005B}.  The luminosity required to explain ultrahigh-energy cosmic rays (UHECR) depends on the assumed spectrum, which is often parameterized by a broken power law with a break at some value $E_c$ (unknown a priori).  According to \cite{2006PhRvD..74d3005B}, the AGN luminosities in cosmic-ray protons needed to account for the UHECR data are $5.6\times 10^{43}$~erg/s, $2.5\times 10^{44}$~erg/s, and $1.1\times 10^{45}$~erg/s, for $E_c=10^{18}$~eV, $E_c=10^{17}$~eV, and $E_c=10^{16}$~eV, respectively.  These estimates are in general agreement with our results.  We also note that, because of the selection effects, the sources observed from large distances are by no means average: they are the brightest AGNs, which generate exceptional power in cosmic rays. 

The spectra of observed gamma rays generated in this fashion depend on the intervening intergalactic magnetic fields.  
It is easy to understand some qualitative features of this dependence in terms of a simplified random-walk description of the proton propagation. 
Let us consider a short pulse of protons emitted from a source at distance $d$.  At later times, the proton pulse broadens 
and takes the shape $f(t,r)$.   The explicit form of $f(t,r)$ was computed by~\cite{2010PhRvD..82d3002A} and will be discussed below. (See also \cite{1978ApJ...222..456A}; \cite{1972MNRAS.157...55W}.) 

At every point in its trajectory, the proton interactions with the cosmic background generate a flux of gamma rays, which quickly (on a kpc length scale) cascade down to energies below the threshold.  From that point on, gamma rays travel without further time delays.  However, during the cascade development, the IGMFs cause some delays (which are longer than the delays of the protons in the IGMFs for $E_\gamma$ below 10~TeV). 

Let us consider the proton propagation in IGMFs.  We assume that IGMFs form a lattice with correlation length $l_c$, in which a  
proton with energy $E_p=10^{17} {\rm eV}$ random-walks over a distance $d \sim 1000\, {\rm Mpc} = n \times l_c$, where $n\sim 10^3$.  

The angle between the proton momentum and the line of sight performs a two-dimensional random walk in small steps of $\sim 10^{-6}$.  When the proton interacts with a background photon, it emits a narrow shower in the direction of the proton's momentum.  The prompt gamma rays are emitted into a narrow angle $(\sim E_p/ {\rm MeV})^{-1}$ for Bethe--Heitler pair production, or $(\sim E_p/ 0.2\, {\rm GeV})^{-1}$ for pion photoproduction.  The cascade develops and broadens this angle, with larger angles at lower energies. To shower in the direction of observer after $n$ steps of random walk, the proton angle should return to 0.  For a 2D random walk, the probability of {\em not} returning to zero 
after $n$ steps is $ \gamma_2(n)= \frac{\pi}{\ln n} + O \left( \frac{1}{(\ln n)^2} \right)$. 
This probability drops below 1/2 for $n > \exp (2 \pi) \sim 5\times 10^2$.   For  $d \sim 1000\, {\rm Mpc}$, $n\sim 10^3$, and so each proton angle returns to the origin about $\sim 1$ time per distance 
traveled. Therefore, the diffusion approximation is justified, and a ``typical'' delay can be computed using the distance traveled,
assuming the random walk. 

Deflection of a proton in a single cell is $\theta_0$.  This deflection and the time delay are determined by the Larmor radius 
$$
R_B= \frac{E}{eB} = 10^5 {\rm Mpc} \ E_{p,17}/B_{-15}, 
$$ 
where $ E_{p,17}$ is the proton energy in units of $10^{17}$~eV, and $B_{-15}$ is the value of the magnetic field in femtogauss. 

Therefore, 
$$
\theta_0= \frac{l_c}{R_B} = 10^{-5} \left(\frac{l_c}{\rm Mpc} \right) \ B_{-15}/E_{p,17}.
$$

Time delay in crossing a single cell is 
$$
\Delta t_0 = \frac{l_c}{c} \theta_0^2 = 10^4 {\rm s} \left(\frac{B_{-15}}{E_{p,17}}\right)^2.
$$

After $n\sim 10^3$ steps of random walk, the time delay is 
\begin{equation}
\tau_p = n  \Delta t_0 \, \sim 10^7 {\rm s}  \left(\frac{B_{-15}}{E_{p,17}}\right)^2.
\label{eqn:proton_delay_E_p}
\end{equation}

Significant time delays are also incurred in the EM showering process. Each observed gamma ray was at some point an electron in the cascade.  The time delay of each gamma ray with an observed energy $E_\gamma$ is dominated by the delay during the lowest-energy ``electron'' phase of this gamma ray.  The electron energy is related to the energy of observed IC $\gamma$ ray by $E_\gamma \propto E_e^2$.  IGMFs act on the electron over a distance of the order of its cooling distance $ D_e \propto 1/E_e$.  The time delay incurred in this process is proportionate to the sum \citep{2008ApJ...682..127I,2008ApJ...686L..67M}  of $D_e$ and the mean free path to pair production $\lambda_{\rm PP}$, which has the same energy dependence (with a much larger prefactor),  
$\lambda_{\rm PP} \propto 1/E_e$.  The resulting delay is 
$$
\tau_e= (\lambda_{\rm PP}+D_e)\theta_e^2/c, \ {\rm where} \ \theta=D_e \frac{ e B}{E_e}.
$$ 
Therefore,
\begin{equation}
\tau_\gamma \approx \tau_e = D_e^2 (\lambda_{\rm PP}+D_e) \frac{ e^2 B^2}{c E_e^2} \propto  \frac{B^2}{E_e^5} \propto  \frac{B^2}{E_\gamma^{5/2}} 
\label{eqn:gamdelay}
\end{equation}
where we have assumed $D_e\sim {\rm kpc}\ll l_c$.  

The proton delay~(\ref{eqn:proton_delay_E_p}) is 
\begin{equation}
\tau_p \propto \frac{B^2}{E_p^2}.
\label{eqn:proton_delay}
\end{equation}

The total time delay of an observed photon is the sum of $\tau_p$ and $\tau_\gamma$: 
\begin{equation}
\tau_{\rm tot} = \tau_p +\tau_\gamma = C_1  \frac{B^2}{E_p^2} + C_2 \frac{B^2}{E_\gamma^{5/2}}, 
\label{eqn:total_delay}
\end{equation}
where $C_1$ and $C_2$ are some constants.  

One can, therefore, expect the following structure of time delays.  The shortest delay time is determined by the delay in the arrival of the highest-energy proton; this time delay is given by Eq.~(\ref{eqn:proton_delay}).  High-energy protons travel faster than the gamma-ray cascades, and they are followed by a tail of trailing gamma rays.  
There are two contributions to the total delay time, which have a different dependence on energy~(\ref{eqn:total_delay}). 
As the second term in Eq.~(\ref{eqn:total_delay}) diminishes with energy, the time delay approaches a plateau independent of the photon energy.  The height of this plateau is 
determined by the energy of the proton.  This agrees with the results of numerical caclculations presented in Figure~\ref{fig1}.

\section{Semi-analytical description}

Let us consider the time delays due to the propagation of protons.  
Protons interactions with extragalactic background radiation (both
CMB and EBL) occur with a very low probability for energies 
below the pion production threshold.   In the pair production process, a proton loses only $\sim 10^{-3}$ of its 
energy in each collision. Thus, one can neglect the energy losses for protons in making some basic estimates. (However, in our numerical 
calculations, we take into account all the energy losses, including adiabatic losses.)   Also, the deflection angles 
can be assumed small for the relevant range of parameters.   
Calculation of the electromagnetic cascade initiated by the secondary gamma rays produced in 
proton-photon interactions  is much more difficult, and there is no simple
analytical approach that could allow one to calculate the distribution function of gamma rays.
Therefore we will employ a hybrid approach by combining an analytical
treatment of protons with a Monte Carlo simulation for the electromagnetic cascade.

To calculate the distribution function of protons, let us consider 
a mono-energetic beam of protons emitted with energy $E$
at some point in time. Random deflections in weak IGMFs result in 
arrival time distribution which is convenient to consider as a function of
time delay parameter $\tau=t-r/c$, where $r$ is the distance to the source
and $c$ is the speed of light. In a small-angle approximation, one can express
the distribution function as follows \citep{2010PhRvD..82d3002A}:
\begin{equation}
f_{A}(E,\tau,r)=\frac{1}{\tau}\label{eqfa}
\left(\frac{c\tau}{r^2\ts} \tilde f_{A}\left(\frac{c\tau}{r^2\ts}\right)\right),
\end{equation}
where
\begin{equation}
\tilde f_{A}(y)=4\pi^2\sum_{n=1}^{\infty}(-1)^{n-1}n^2 e^{-2\pi^2n^2y}
\end{equation}
\begin{equation}
\ts=\frac{l_c}{5}\left(\frac{e}{E}\right)^2 \langle B^2\rangle.
\end{equation}
Here $\ts$ is the mean square deflection angle per unit length. The correlation
length $l_c$ and the mean square of magnetic field $\langle B^2\rangle$
enter in Eq.~(\ref{eqfa}) as parameters. The normalization of the function
$f_{A}$ is so that $\int \limits_0^{\infty}f_{A}d\tau=1$. Then the distribution function
of protons injected with energy spectrum $J_p(E)$ at the distance $r$ from
the source is 
\begin{equation}
f_p(E,\tau,r)=\frac{J_p(E)f_{A}(E,\tau,r)}{r^2}.
\end{equation}
Let protons interact with low energy photon field $f_{ph}(\epsilon)$. The
protons with a monoenergetic distribution (normalized to one particle) produce
electron-positron pairs at the rate $\Phi(E_e,E_p)$, where $E_p$ is the
energy of protons, $E_e$ is the energy of pairs. Following \cite{Kelner}, one
can express $\Phi(E_e,E_p)$ as follows:
\begin{equation}
\Phi(E_e,E_p)=c^2\int d\epsilon \frac{d\Omega}{4\pi}
f_{ph}(\epsilon)\frac{k\cdot u_p}{\epsilon \gamma_p}
\int\delta(E_e-c(u_{lf}\cdot p_e))d\sigma\,,
\end{equation}
where $k$,\,\,$u_p$, and $u_{lf}$ are four-velocities of photon, proton,
and the laboratory frame, respectively; and $p_e$ is four-momentum of electron
(or positron). $\gamma_e$ and $\epsilon$ are the proton Lorentz factor and
the energy of photon in the laboratory frame. $d\sigma$ is the Bethe-Heitler
cross section. The photon field used in this calculations includes the CMB
and EBL, and one can neglect the redshift evolution. 

Then distribution function of electrons
produced at the distance $r$ with inherited time delay $\tau$ is
\begin{equation}\label{del}
f_{e}(E_e,\tau,r)=\int dE_p\frac{J_p(E_{p})f_{A}(E_p,\tau,r)}{r^2}\Phi(E_e,E_p).
\end{equation}
These electrons initiate an electromagnetic cascade. Let  
$f_{cas}(E_e,E_{\gamma},s)$ be the number of photons with energy $E_{\gamma}$
produced in the cascade initiated by an electron with energy $E_e$ at the
distance $s$ from the observer and detected at the point of observation.
The mean time delay of the photons is $\tau_{cas}(E_e,E_{\gamma},s)$. The
extragalactic magnetic field is a parameter. Then for the UHE proton source
at the distance $d$ the number of photons produced in the cascade with time
delay  $\tau=\tau_{cas}+\tau_{prot}$ is  
\begin{equation}\label{fend}
f_{\gamma}(E_{\gamma},\tau,d)=
\int\limits_{0}^{d}dr \int dE_{e}
f_{e}(E_e,\tau-\tau_{cas}(E_e,E_{\gamma},r),d-r)f_{cas}(E_e,E_{\gamma},r)
\end{equation}
The Eqs.~(\ref{eqfa}), (\ref{del}) and (\ref{fend}) constitute the integral
which is calculated numerically. The functions $f_{cas}$ and $\tau_{cas}$ actually
depend on redshift but not on distance, therefore we use the relation
\begin{equation}
dr=\frac{c}{H_0}\frac{1}{(1+z)\sqrt{(1+z)^3\Omega_m+\Omega_{\Lambda}}} dz,
\end{equation}
to express distance via redshift and perform integration over $z$.

\begin{figure}
\plotone{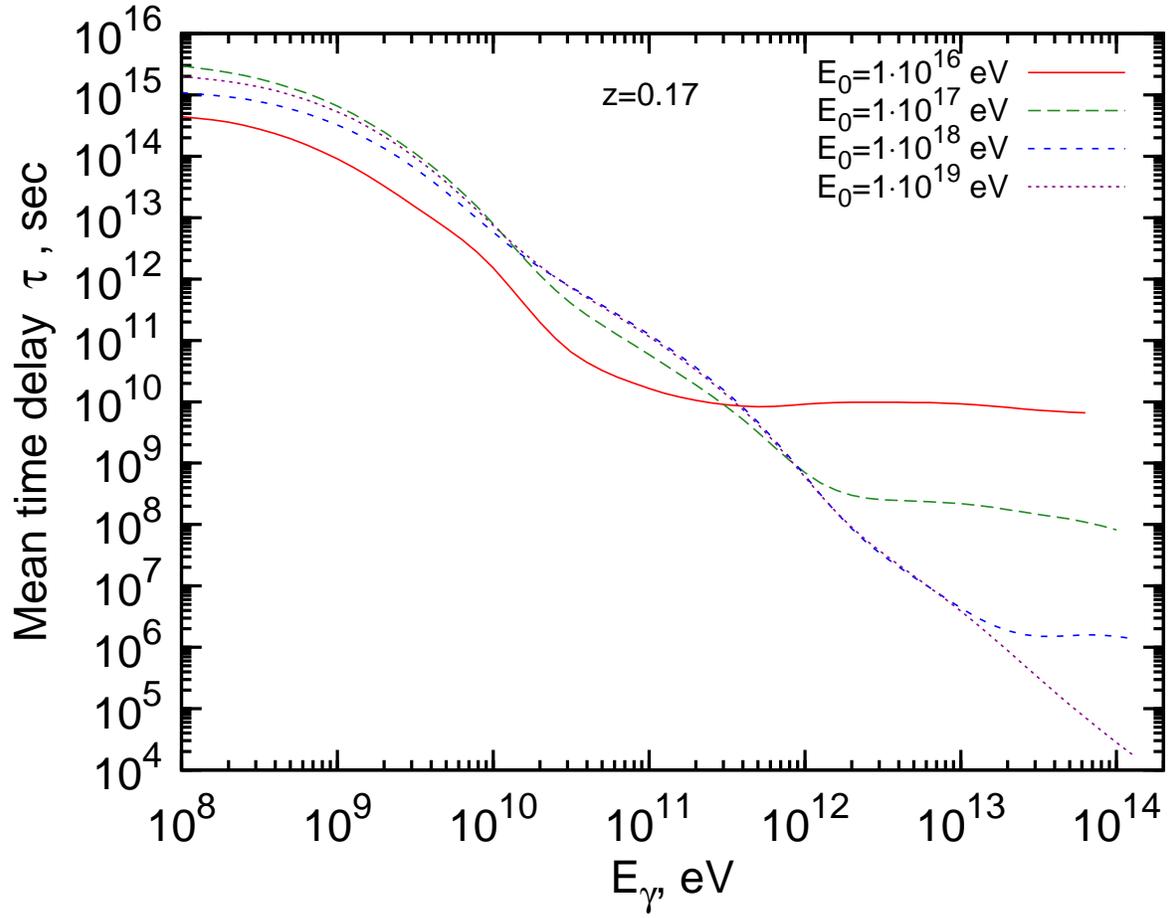}
\caption{\label{fig1}  Mean time delay of gamma rays at redshift $z=0.17$
for different cutoff energies $E_{0}$ of proton spectrum.  
}
\end{figure}

\begin{figure}
\plottwo{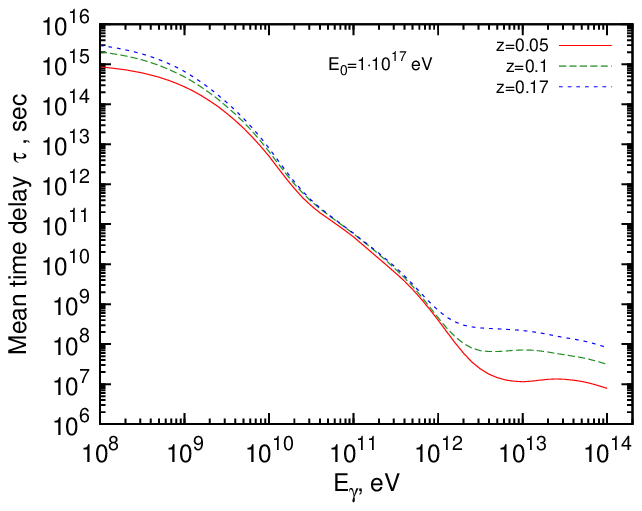}{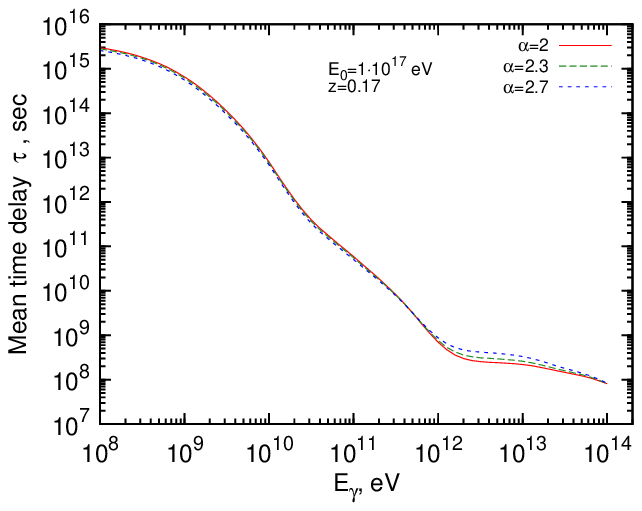}
\caption{\label{fig2}Left panel: mean time delay of gamma rays for the sources
at different redshifts and the proton spectrum with cutoff energy $E_{0}=10^{17}$
eV. Right panel: mean time delay of gamma rays for the source
at $z=0.17$ and the proton spectrum with cutoff energy $E_{0}=10^{17}$
eV and different spectral indices $\alpha$.  
}
\end{figure}

We assume that the source produces a power-law spectrum of protons with a spectral index $\alpha$
and the  energy range from $0.1 E_0$ to $E_0$. The results for the mean time delay of gamma rays
are presented in Figs.~\ref{fig1} and \ref{fig2} as a function of the varying cut-off energy, distance to the source
and spectral index.  In these calculations we assumed that the intergalactic magnetic
field has the strength $B=10^{-15}$~G and the coherence length $l_c=1$~Mpc. Unless specified otherwise, we use $E_0=10^{17}$ eV, $\alpha=2$.

\begin{figure}
\begin{center}
\begin{tabular}{cc}
\includegraphics[width=0.5\textwidth,angle=0]{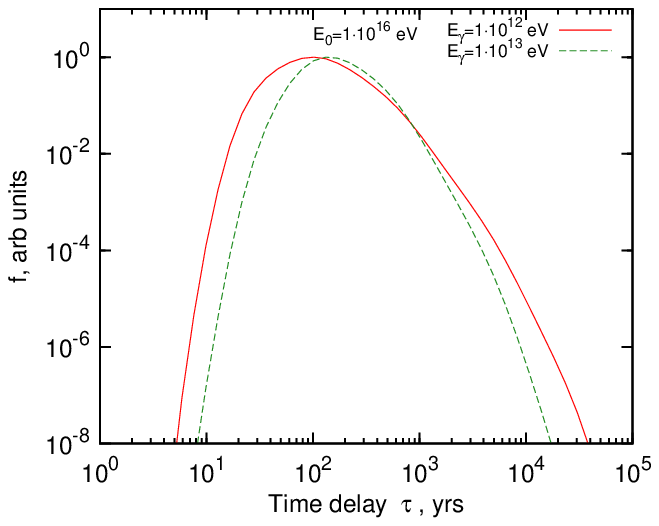}& 
\includegraphics[width=0.5\textwidth,angle=0]{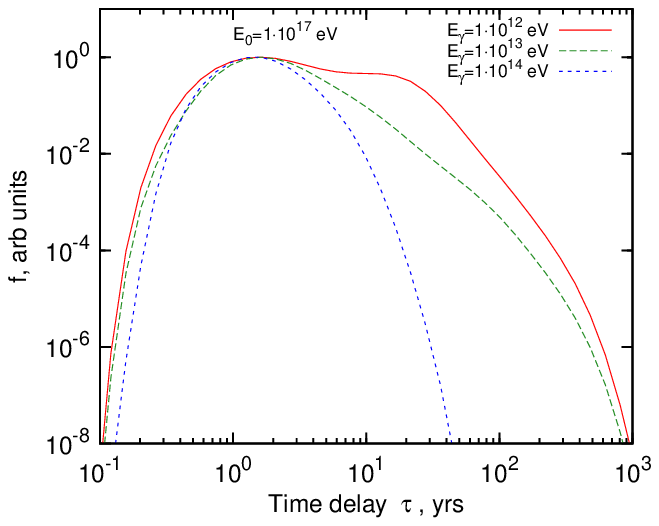}\\
\includegraphics[width=0.5\textwidth,angle=0]{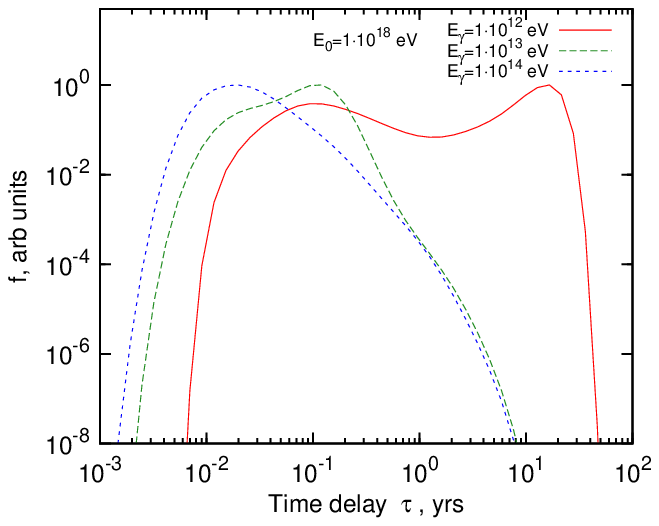}&
\includegraphics[width=0.5\textwidth,angle=0]{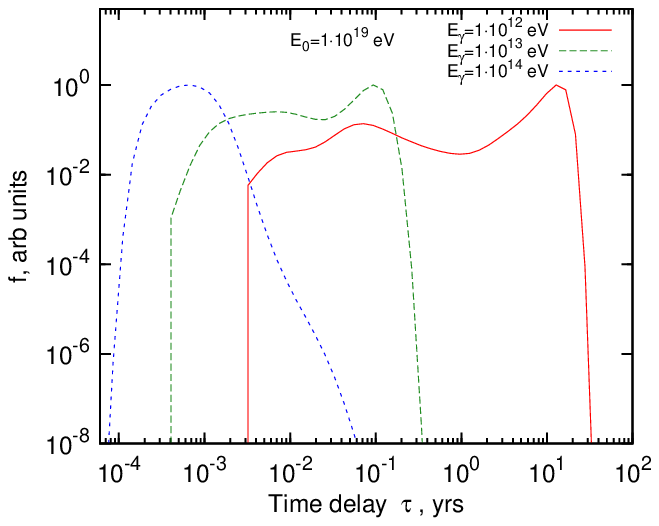}\\
\end{tabular}
\caption{\label{fig3} Time delay distribution of gamma rays in arbitrary 
units (the maximum of distribution is normalized to unity) at different
energies from the source at $z=0.17$. Each plot corresponds to the proton
spectrum with different cutoff energy $E_{0}$ and spectral
index $\alpha=2$. The injected spectra of protons are taken in the range
from $0.1 E_0$ to $E_0$.}
\end{center}
\end{figure}

\begin{figure}
\begin{center}
\includegraphics[width=0.8\textwidth,angle=0]{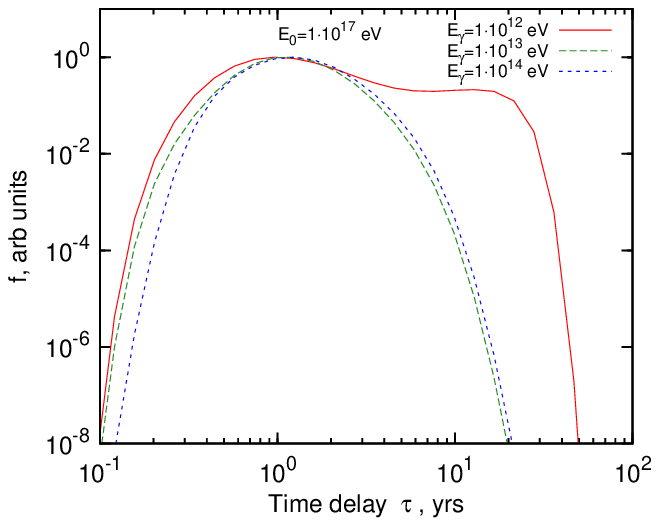}
\caption{\label{fig4}  Time delay distribution of gamma-ray in arbitrary
units (the maximum of distribution is normalized to unity)  from the
source at $z=0.17$. The injected spectrum of protons is almost monoenergetic
with energy $E=10^{17}$ eV. }
\end{center}
\end{figure}

\begin{figure}
\begin{center}
\includegraphics[width=\textwidth]{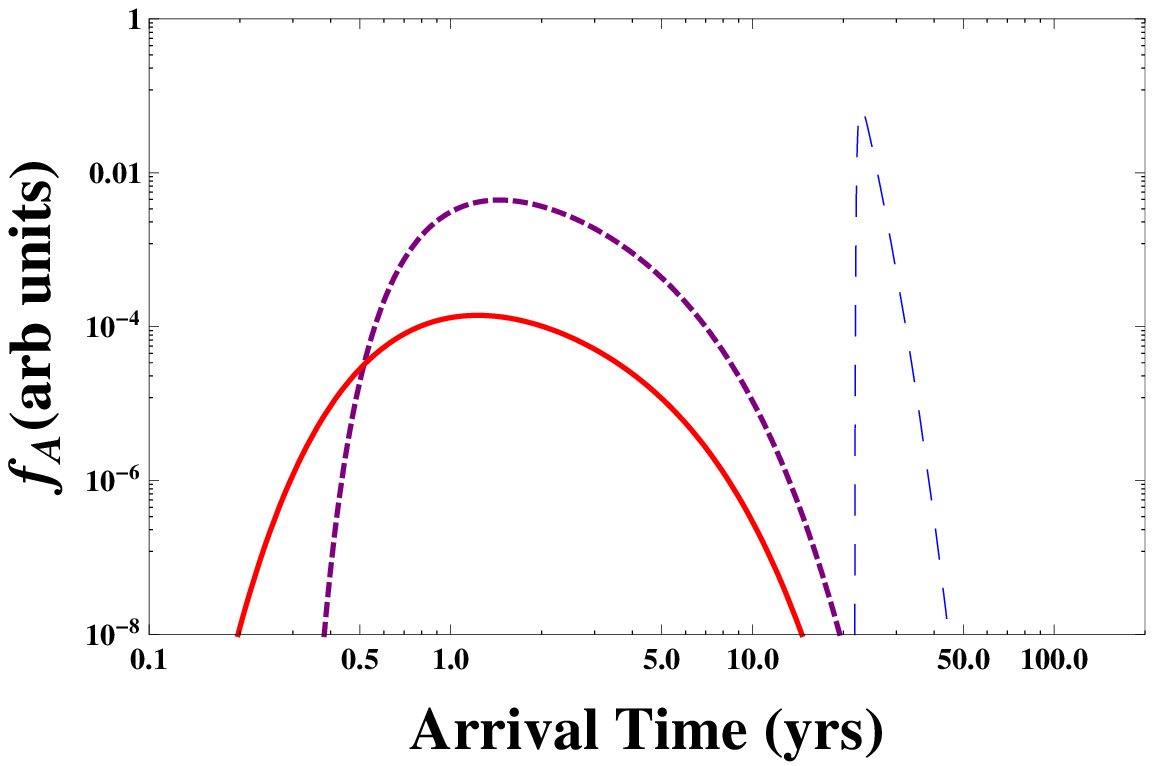} 
\hspace{0.5cm}
\caption{
Arrival time probability distribution in arbitrary units for secondary gamma rays with energies 1~TeV (blue, long-dashed line), 10~TeV (purple, short-dashed line) and 100~TeV (red, solid line). 
Results are shown for a cosmic ray source at $z=0.2$ with a high energy cutoff of $10^8$~GeV and an IGMF of $10^{-15}$~G.\label{gam1_100TeV_1e15}}
\end{center}
\end{figure}

The advantage of the analytical description presented above is the possibility
to study the time delay distribution of gamma rays for a variety of initial proton spectrum
parameters. The numerical Monte Carlo approach described below has computational limitations on the number
of initial particles, which can complicate the study of how a proton spectrum
with a wide energy range can affect the time delays of gamma rays. The protons injected
with slightly different energies would have the time distribution which is
similar to the time distribution for a monoenergetic proton beam. In contrast, 
the time distribution of protons with a broad energy spectrum is a sum of
time distributions of protons with different energies, which is stretched
out in time. This would spread the arrival times of gamma rays along a large
time span. The illustration of this effect can be seen from the comparison
of the second panel of Fig.~\ref{fig3} and Fig.~\ref{fig4}. 
The range of the proton energies has a strong effect on the lower energy gamma rays, 
whereas the time distribution for $E_{\gamma}=1\cdot10^{14}$~eV  does not change significantly.
This is because, in the case of a broad spectrum, protons with different energies can contribute 
gamma rays of a given energy.  On the other hand, only the protons of highest energies are responsible for the production
of gamma rays of $E_{\gamma}=1\cdot10^{14}$~eV.  Therefore, the corresponding time distribution 
is similar to the one for the monoenergetic protons ({\em cf.} Fig.~\ref{gam1_100TeV_1e15}).
 
The flux of gamma rays arriving at any given time comprises contributions from protons 
at different points along the line of sight.  We obtain this flux by integrating over the proton 
distributions shifted by a time delay incurred in the electromagnetic cascade.  The latter delays were obtained 
from numerical calculations using a single Monte Carlo numerical run. The results are shown in Figs.~\ref{fig3} and Fig.~\ref{fig4}. 
The multiple-peak structure apparent in these curves is the result of adding contributions from different distances 
with the delay profile obtained from a single numerical run.  If we adopted a different approach and used an averaged delay 
profile, as in Fig.~\ref{fig:numerical_runs}, the ``many-peak`` structure would be erased.

\section{Numerical Monte-Carlo calculations}

In addition to the semi-analytical results we also performed a full scale Monte Carlo simulation to track the arrival times of individual particles. The source was modeled by an instantaneous pulse of protons to represent the Green's function needed to calculate the distribution of arrival times. Particles are advanced in time steps of roughly $0.1-1$~kpc, updating momentum, position and time delay with all relevant interactions taken into account. Gamma rays arriving at the $z=0$ surface are binned and the mean arrival time and standard deviation are calculated.

The proton energy loss processes are well studied~\citep{1994APh.....2..375S} and can be described by a standard approach. 
We calculate all the relevant energy losses, including adiabatic losses and the losses due to the interactions with photon backgrounds. 
The most important contributions to secondary photon production  are photopion production and proton pair production (PPP).

The photopion production processes involve the following reactions: 
\begin{eqnarray}
p+\gamma_b & \rightarrow & n + \pi^+ \nonumber \\
p+\gamma_b & \rightarrow & p + \pi^0
\label{eqn:pion}
\end{eqnarray}
where $\gamma_b$ is either a CMB or EBL photon.  PPP occurs in the reaction 
\begin{equation}
p+\gamma_b \rightarrow p + e^+ +e^-.
\label{eqn:PPP}
\end{equation}
The pair production on the CMB is the dominant reaction, but pion photoproduction on EBL also contributes.  
Pion photoproduction on CMB has a threshold above $10^{19}$~eV, but pion production on EBL is possible for all energies we consider.   
The efficiency of energy transfer to the electromagnetic shower depends on the proton energy and on the distance to the source. 
A more detailed discussion is presented elsewhere~\citep{2012arXiv1206.6715A}.

The mean interaction length, $\lambda$, for a proton of energy E traveling through a photon field is given by
\begin{equation}
[\lambda[E]]^{-1}=\frac{1}{8\beta E^2}\int^\infty_{\epsilon_{\rm min}}\frac{n(\epsilon)}{\epsilon^2}
\int^{s_{\rm max}(\epsilon,E)}_{s_{\rm min}}\sigma(s)(s-m_p^2)dsd\epsilon, 
\end{equation}
where $n(\epsilon)$ is the differential photon number density of photons of energy $\epsilon$, and $\sigma(s)$ is the appropriate total cross section for the given process for the center of momentum (CM) frame energy squared, $s$, given by
\begin{equation}
s=m_p^2+2\epsilon E(1-\beta \cos\theta),
\end{equation}
where $\theta$ is the angle between the proton and photon, and $\beta$ is the proton's velocity.\\
For pion photoproduction,
\begin{equation}
s_{\rm min}=(m_p^2+m_{\pi}^2)^2
\end{equation}
and
\begin{equation}
\epsilon_{\rm min}=\frac{m_{\pi}(m_{\pi}+2m_p)}{2E(1+\beta)}.
\end{equation}
For proton pair production 
\begin{equation}
s_{\rm min}=(m_p^2+2m_e^2)^2
\end{equation}
and
\begin{equation}
\epsilon_{\rm min}\approx \frac{m_e(m_e+m_p)}{E}.
\end{equation}
For both processes,
\begin{equation}
s_{\rm max}(\epsilon,E)=m_p^2+2\epsilon E(1+\beta).
\end{equation}
Both pions and neutrons quickly decay via the processes
\begin{eqnarray}
n & \rightarrow & p + e^- + \bar{\nu}_e,\nonumber \\
\pi^+ & \rightarrow & \mu^+ + \nu_{\mu} \rightarrow e^+ + \nu_e + \bar{\nu}_\mu + \nu_\mu, \nonumber \\
\pi^0 & \rightarrow & 2\gamma.
\end{eqnarray}
The outgoing distribution functions for pion photoproduction were generated using the SOPHIA package~\cite{2000CoPhC.124..290M}.

Primary gamma rays and gamma rays produced from the above equations can interact and pair-produce on background photons. The resulting electron positron pairs will IC scatter CMB photons. The upscattered photons can once again pair produce, this chain reaction is known as electromagnetic (EM) showering.\\
The interaction length for photons for pair production off the EBL is
\begin{equation}
[\lambda]^{-1}=\Big(\frac{m_e^2}{E}\Big)^2\int^\infty_{\frac{m_e^2}{E}}\epsilon^{-2}n(\epsilon)\int^{\frac{\epsilon E}{m_e^2}}_1 2s\sigma(s)dsd\epsilon,
\end{equation}
where
\begin{equation}
\sigma = \frac{1}{2}\pi\Big(\frac{e^2}{m_e^2}\Big)^2(1-\beta^2)\Big[(3-\beta^4) \ln\frac{1+\beta}{1-\beta}-2\beta(2-\beta^2)\Big]
\end{equation}
and
\begin{equation}
\beta=(1-1/s)^{1/2},
\end{equation}
and $n(\epsilon)$ is the differential photon number density of photons of energy $\epsilon$.

To simulate magnetic field effects, the IGMF is modeled by cubic cells of a given magnetic field strength with sides equal to the chosen correlation length, $l_c$, and a random direction. Particles are moved forward in fine time steps and the deflection of the particle is calculated using the Larmor radius and IGMF direction. Time delays for charged particles are calculated in comparison to a photon traveling in a straight line to the observer.

\begin{figure}
\begin{center}
\includegraphics[width=\textwidth]{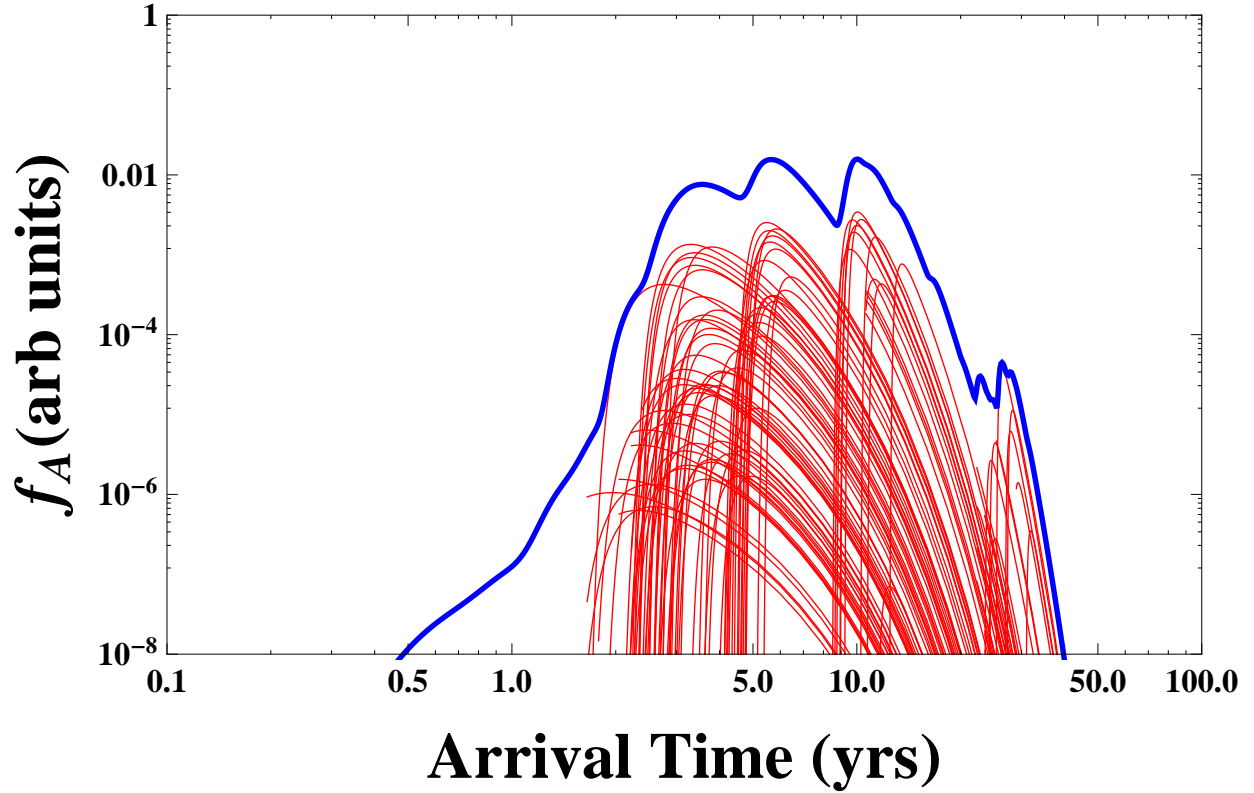}
\hspace{0.5cm}
\caption{\label{fig:numerical_runs}
Arrival time probability distribution in arbitrary units for 1 TeV secondary photons for multiple numerical runs. The results shown are for roughly 300,000 secondary photons with an IGMF of $10^{-15}$~G and UHECR cutoff of $10^{10}$~GeV. The blue thick line represents the sum of all distributions and the thin red lines are a representative set of distributions.}
\end{center}
\end{figure}

For the analysis of time delays, we have performed multiple runs and averaged the results, as shown in Fig.~\ref{fig:numerical_runs}.  

\begin{figure}[ht]
\begin{center}
\includegraphics[width=\textwidth]{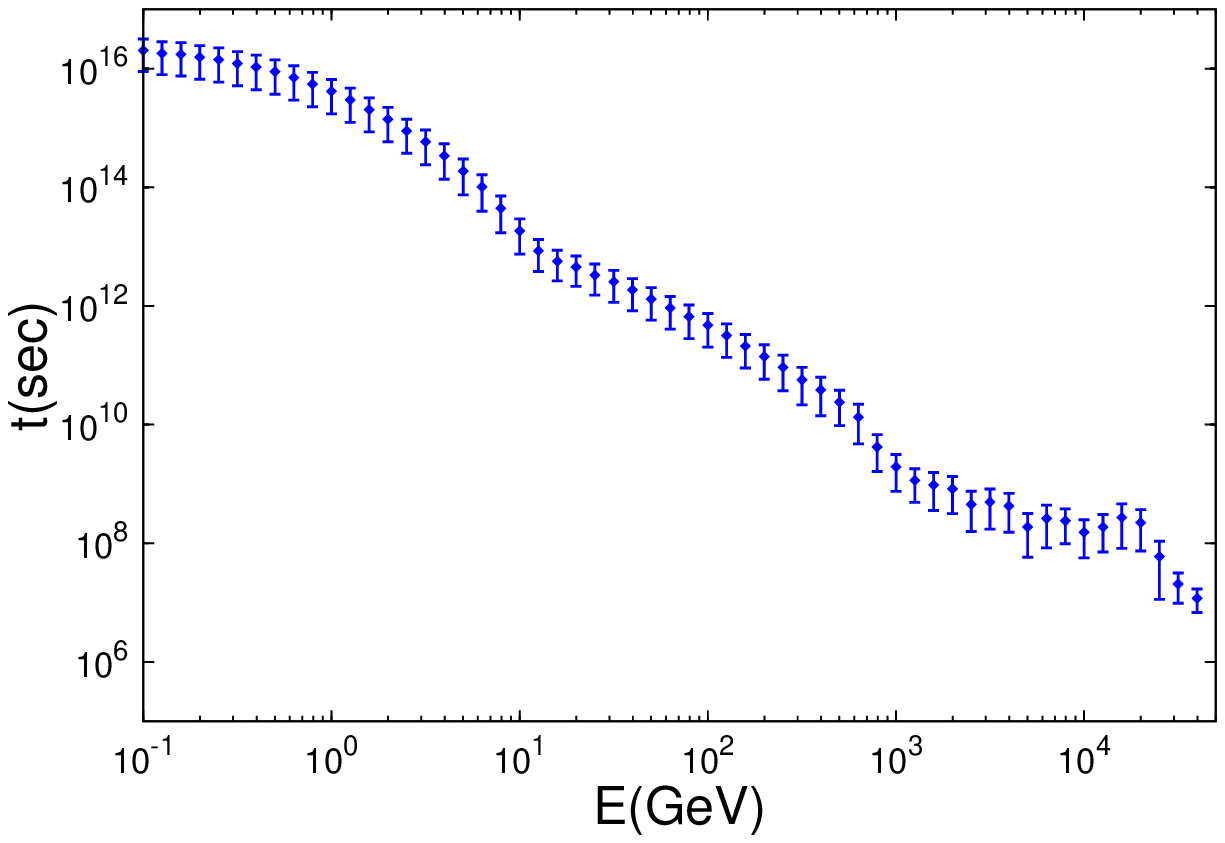}
\hspace{0.5cm}
\caption{Arrival time delays from an instantaneous pulse emitted by a source at $z=0.2$, assuming $B=10^{-15}$~G with $l_c=1$~Mpc correlation length. 
\label{Greensfunction15approx}}
\end{center}
\end{figure}

The results of the simulation are shown in Fig.~\ref{Greensfunction15approx}, where  delays from deflections in the IGMF are shown for a source at $z=0.2$.  It is evident that secondary photons produced at large distances conform to the power-law behavior as in 
Eq.~(\ref{eqn:gamdelay}). This approximate power law is illustrated in Fig.~\ref{Greensfunction15approx} by a dashed line. The flattening at low energies can be understood by the way the code handles deflections. Particles are moved forward in time steps of roughly $0.1-1$~kpc and deflections are assumed to be less than $\pi$ within a single time step. For the lowest energies  this is not always true and the code will underestimate the deflection, thus producing time delays below the power law. 

\section{Discussion and conclusion}

The main qualitative features of the Green's function computed numerically and shown in Fig.~\ref{fig1} and Fig.~\ref{Greensfunction15approx}
can be easily understood. For lower energies (below TeV), time delays $\tau \propto B^2 E^{-5/2} d$, where $d$ is distance to the source. The nearby showers arrive before distant showers, so that the late arriving gamma rays have lower energies and longer delays. The plateau that develops at $E>1$TeV is due to the prompt showers emitted by the protons nearby, for which the time delays are determined by the proton deflections in IGMFs. In the absence of cosmic rays, the spectrum would drop above 1~TeV, and the multi-TeV gamma rays would not be observed. 
 
For sub-TeV secondary gamma rays the electromagnetic cascade delays are always longer than the proton delays, and the arrival photons peak at the time given by Eq.~(\ref{eqn:gamdelay}). At energies above  TeV, the proton delays come to dominate, in accordance with the broken power law in Eq.~(\ref{eqn:total_delay}).  The numerical results differ somewhat from the scaling in Eq.~(\ref{eqn:total_delay}).  In particular, at low energies, 
the delays appear to scale as $E^{-2}$ rather than $E^{-2.5}$.  The difference can be explained by a combination of several effects. For electron energies below 30 GeV, the cooling distance exceeds the magnetic field correlation length, which we assumed to be $l_c\sim 1$~Mpc.  This changes the 
energy dependence in Eq.~(\ref{eqn:gamdelay}) because the energy-dependent cooling distance must be replaced by the constant correlation length.  Furthermore, integration over energies in the cascade affects the power-law behavior.  For these reasons, our basic estimates in section 2 were not expected to capture all the features evident in the numerical results. 

\begin{figure}
\begin{center}
\includegraphics[width=\textwidth]{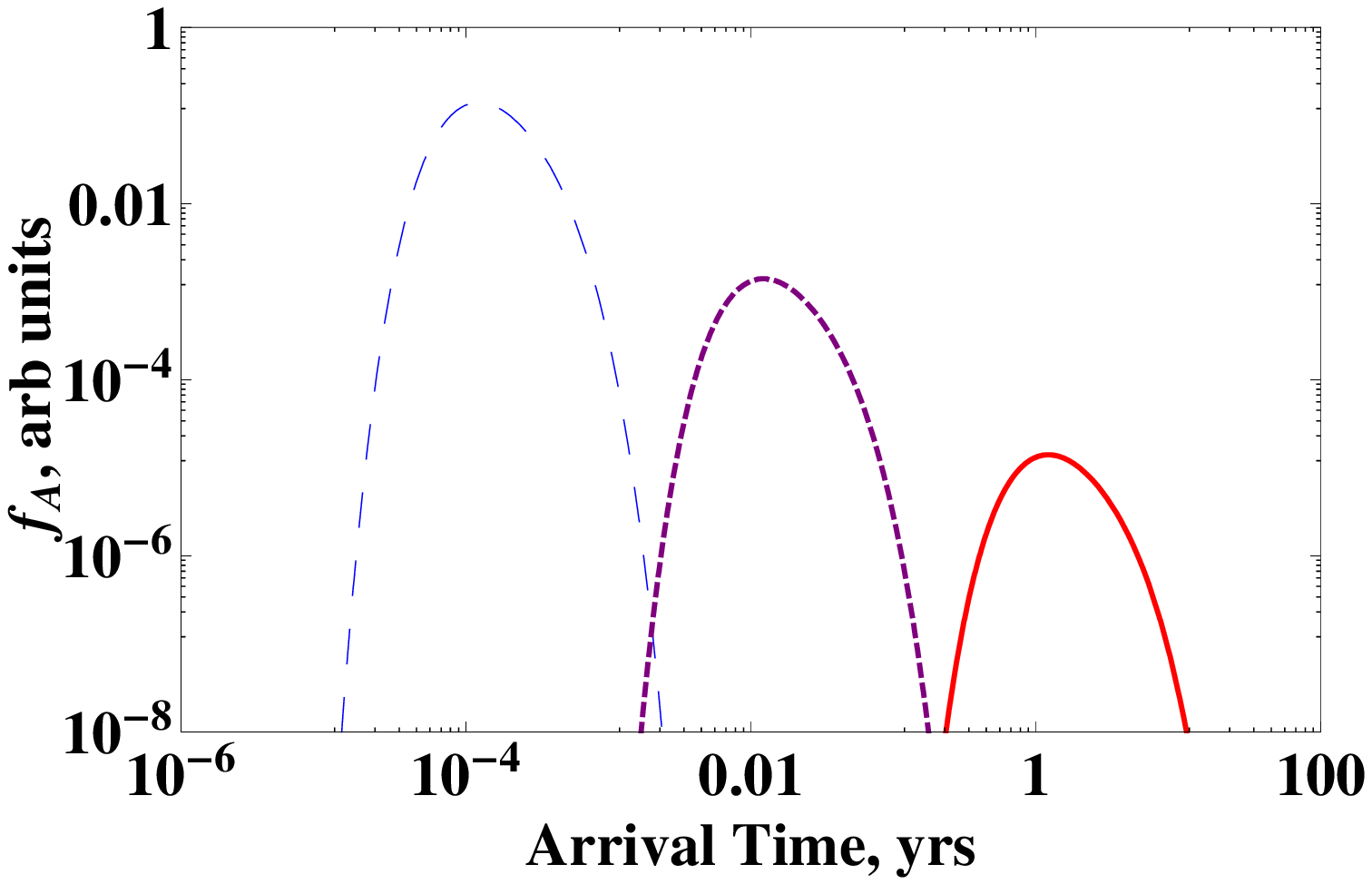}
\hspace{0.5cm}
\caption{
Arrival time probability distribution in arbitrary units for primary cosmic rays. Results are shown for a cosmic ray source at $z=0.2$ with a high energy cutoff of $10^{10}$~GeV (blue, long dashed), $10^9$~GeV (purple, short dashed) and $10^{8}$~GeV (red, solid) and an IGMF of $10^{-15}$~G.\label{pro1e15}}
\end{center}
\end{figure}

The proton delay is strongly dependent on the high energy cutoff of the cosmic ray source, which affects the energy at which the proton delays begin to dominate.  This can be seen in Fig.~\ref{pro1e15}. This behavior is further illustrated in Fig.~\ref{fig3} and Fig.~\ref{gam1_100TeV_1e15}, where one can see that the proton delays begin to dominate at $E\sim10$~TeV for a proton high energy cutoff of $10^8$~GeV and an IGMF~$=10^{-15}$~G. 

The distribution of gamma-ray arrival times depends on the injection spectrum of protons, as one can see from a comparison of Fig.~\ref{fig3} and Fig.~\ref{fig4}.  
This is in contrast with the {\em spectra} of gamma rays, which are {\em not} sensitive to the proton injection spectrum \citep{Essey:2009ju,Essey:2010er}. Hence, one can, at least in principle, 
learn about the proton injection spectrum from timing observations, but not from the spectra alone.  (Neutrino spectra also depend on the proton injection spectrum \citep{Essey:2009ju}.)
Furthermore, stochastic broadening illustrated in Fig.~\ref{fig:numerical_runs} also affects the predictions.

Based on our results, the observed time variability should be washed out on time scales shorter than $\sim 0.1$~yr, for distant blazars ($z>0.2$), at TeV and higher energies.  Time variability can be present for $z>0.2, E>1$~TeV on the time scales of $0.1-1$~yr.  If gamma rays with $E\sim 10^2$~TeV are observed, they can exhibit variability on shorter time scales. 

Of course, one must also consider the delays the cosmic rays undergo at the source. Blazars are known to be highly variable, and this variability could affect the shape of the observed spectrum. The magnetic fields within galaxies are on the order of $1~\mu G$ which can lead to significant delays in the source.  On the other hand, the structure of magnetic fields in front of the blazar jets is not known.
Furthermore, the effect of the source variability would be to suppress the observed power of the source by a factor 
 \begin{equation}
 f_{\rm damp}\sim N_{\rm active} \left ( \frac{t_{\rm active}}{t_{\rm delay}} \right ), 
 \end{equation}
where $t_{\rm delay}$ is the typical proton delay at the source, $t_{\rm active}$ is the typical time the source is active or flaring and $N_{\rm active}$ is the number of times the source is active in the time period $t_{\rm delay}$. This damping should not be a significant effect \citep{Dermer:2010mm}, especially since the typical deflections at the source are not big enough to affect the beaming factors assumed in popular models.

An alternative situation is that the magnetic fields within the blazar jet are not randomly distributed, but are, instead, strongly correlated with the direction of the jet. Blazar jets emit an extremely large amount of charged matter and the wind in the direction of the jet can eliminate any random-field configuration that one usually expects in a galaxy. 
Thus, it is possible that cosmic rays  escape the source along the jet with very small time delays, preserving the intrinsic variability of the source. In this case, delays in the intergalactic medium can broaden the intrinsic variability to the energy dependent timescales of these delays.
 
At energies where the optical depth of the observed gamma rays is below one, we expect the signal to be primary gamma rays, and any variability in this signal is indicative of the source variability. This variability should not depend strongly on the energies of the gamma rays, but rather on the scale of the structure at the source producing the gamma rays. 
 
However, we expect a very different behavior for energies at which primary gamma rays are significantly attenuated by pair production off EBL. In the case of strongly correlated magnetic fields in the jet, we expect that the variability should show different structure in the low energy component, where it should depend on the energy. The spectrum should show variability on shorter timescales for higher energies until some critical energy $E_c \sim {\rm TeV}$,  where the timescales cease to decrease further, thanks to the domination of the cosmic ray contribution. In the case of large delays within the source, we expect all variability to be washed out at these higher energies (typically around a TeV for most observed sources). 

Some exceptionally bright flares can come through around $E_c$ and rise above the {\em pedestal} created by the stochastic arrival times of protons.  Such flares should have distinctly softer spectra than the hard  pedestal, which can be a means of distinguishing these flares from the stochastic pedestal. 

For most of discussion, we have assumed that IGMFs have strengths are of the order of a femtogauss.  
This range is suggested by the spectral fits to the data \cite{Essey:2010nd}.  However, field strengths well below a femtogauss can be consistent with the data as well.  In the case of 
very weak IGMFs, the time delays become smaller, since $\tau \propto B^2 $.  For $ B\sim 10^{-18}$~G, the time delays can be as short as minutes. 

We have also assumed that the strength of IGMF is constant on average, which is a good assumption for propagation in the voids.  However, if the line of sight intersects a filament of 
stronger, {\em e.g.}, nanogauss magnetic field, the reduction of the secondary photon signal depends on the size of the filament and on its location.  Thin filaments can only intercept as small 
fraction of protons within the 0.1 degree associated with a given source.  However, a thick filament or a sheet of strong field can deflect protons, reducing the secondary signal.  

Temporal structure of gamma-ray signals can be used to measure the IGMF structure and EBL intensity in different directions, on a source-by-source basis. In addition, it may provide a way 
to probe the high energy cutoff of cosmic ray sources,  as well as the spectrum of EBL. A statistical analysis on multiple bins of data is needed to determine the variability at different energies. 
This presents challenges at the highest energies because of the low statistics, but longer observation times and the advent of next generation experiments should make this analysis increasingly powerful.

\section{Acknowledgments} 
The authors thank S.~Ando and K.~Murase for helpful discussions. The work of W.E. and A.K. was supported in part by DOE grant DE-FG03-91ER40662. 
A.K. appreciates the hospitality of the Aspen Center for Physics, which is supported by the NSF grant No. PHY-1066293.



\end{document}